# Compact Ultra-low Loss Optical True Delay Line on Thin Film Lithium Niobate

*Yuan Ren,[†] Boyang Nan†, Rongbo Wu\*, Yong Zheng, Ruixue Liu, Yunpeng Song, Min Wang, and Ya Cheng\**

Yuan Ren, Boyang Nan, Yong Zheng, Ruixue Liu and Prof. Ya Cheng
State Key Laboratory of Precision Spectroscopy, East China Normal University, Shanghai 200062, China

Yuan Ren, Boyang Nan, Rongbo Wu, Yong Zheng, Ruixue Liu, Yunpeng Song, Min Wang and Prof. Ya Cheng
The Extreme Optoelectromechanics Laboratory (XXL), School of Physics and Electronic Science, East China Normal University, Shanghai 200241, China
E-mail: rbwu@phy.ecnu.edu.cn

Prof. Ya Cheng
State Key Laboratory of High Field Laser Physics and CAS Center for Excellence in Ultra-intense Laser Science, Shanghai Institute of Optics and Fine Mechanics (SIOM), Chinese Academy of Sciences (CAS), Shanghai 201800, China
Collaborative Innovation Center of Extreme Optics, Shanxi University, Taiyuan 030006, China.
Collaborative Innovation Center of Light Manipulations and Applications, Shandong Normal University, Jinan 250358, China
Shanghai Research Center for Quantum Sciences, Shanghai 201315, China
Hefei National Laboratory, Hefei 230088, China
E-mail: ya.cheng@siom.ac.cn

**Keywords:** optical true delay line, photonic integrated circuit; thin film lithium niobite

We report the fabrication of an 8-meter-long thin-film lithium niobate (TFLN) optical true delay line (OTDL) using the photolithography-assisted chemomechanical etching (PLACE) technique, showing a low transmission loss of 0.036 dB/cm in the conventional telecom band.

## 1. Introduction

Optical true delay lines (OTDLs) constructed using ultra-low-loss optical fibers are widely used in various applications, including rotation sensing, microwave photonics, and optical coherence tomography. [1–3] Optical true delay lines (OTDLs) constructed using ultra-low-loss optical fibers are widely used in various applications, including rotation sensing, microwave photonics, and optical coherence tomography. [4–9] Optical true delay lines (OTDLs) constructed using ultra-low-loss optical fibers are widely used in various applications, including rotation sensing, microwave photonics, and optical coherence tomography. [10–13] Additionally, TFLN has a wide transparent window with low intrinsic material absorption, which in principle allows for the implementation of low-loss OTDLs with very long lengths. Moreover, TFLN serves as an ideal host material for doping rare-earth ions, facilitating monolithic integration of laser sources and amplifiers. [14,15] However, implementing very long OTDLs on TFLN imposes new challenges on the micro-fabrication techniques, as waveguide loss must be minimized to an unprecedented level and the fabrication process must be scalable to ensure these low-loss characteristics are preserved in tightly folded waveguide structures over a wafer-scale footprint. Whilst ultra-low loss or wafer-scale TFLN-based photonic devices have also been successfully fabricated using electron beam lithography (EBL) [16] and deep ultraviolet (DUV) lithography [17] separately, neither technique can achieve both ultra-low-loss fabrication and large exposure area simultaneously.

To address the challenges, we have been developing and perfecting the photolithography-assisted chemomechanical etching (PLACE) technique since 2018. [18–20] The PLACE technique uses a thin layer of chromium (Cr) coated on TFLN as hard mask which is rapidly patterned using high-precision femtosecond laser direct writing system. The transfer of mask pattern to the underneath TFLN is achieved using a chemomechanical polishing (CMP) process. Since the femtosecond laser direct writing technology allows for continuous writing of arbitrary mask patterns across the entire 12-inch wafer, it facilitates the fabrication of large-scale photonic structures without stitching and in turn avoiding the extra loss caused by the misalignment between two stitched waveguides. The CMP technique ensures that the fabricated

TFLN photonic devices have sub-nanometer sidewall roughness and significantly reduces the optical scattering loss, which is vital for low-loss photonic devices. Combining these two techniques makes the PLACE method particularly effective for manufacturing large-scale photonic devices with ultra-low optical transmission loss. Recently, we fabricated an OTDL with a wavelength length of up to 1 meter using the PLACE technique, achieving a loss as low as 0.03 dB/cm. [9] However, the TFLN waveguide is generated using Ta2O5 as cladding layer, thus the bend radius must be relatively large as Ta2O5 has a low refractive index contrast compared with lithium niobate. To facilitate scaling up the photonic integrated circuits (PICs), nowadays the TFLN waveguides fabricated using PLACE technique are cladded by fused silica to enable small bend radius. The improvement results in the largest integrated arrays of Mach-Zehnder interferometers (MZIs) on TFLN to date on which various AI tasks have been performed. [13]

In this work, we report fabrication of a TFLN OTDL with a length up to 8 meters. We examine the optical loss in the conventional telecom band (e.g., C band), showing an average optical loss of $0.036 \pm 0.002$ dB/cm. In addition, we inject a data stream through the TFLN OTDL to demonstrate its data buffering functionality.

## 2. Methods and Results

The schematic of the OTDL demonstrated in this study is presented in Figure 1 (a). The OTDL is coiled within a 2 cm × 0.5 cm area, comprising 192 turns with a spacing of 15 μm between adjacent turns. The total number of 90° waveguide bends reaches 772 in the fabricated OTDL. To mitigate the losses occurring in the mode conversion from the straight segments to the waveguide bends, as well as to address the radiative losses of the bent waveguide modes, 90° partial Euler waveguide bends are utilized in our design. [21] Figure 1 (b) illustrates the geometry of a 90° partial Euler bend and defines the parameters. The blue regions feature curvature-continuous Euler bends, while the red regions are essentially circular arcs. The bend parameter p indicates the portion of the bend that exhibits a linearly increasing curvature. The effective radius $R_{eff}$ of the partial Euler bends coincide with those of a circular bend of the same radius. The minimum radius $R_{min}$ indicates the radius of curvature for the circular part of the partial Euler bend. Figure 1 (c) presents numerically simulated transmission losses for different modes of TFLN waveguides with different curvatures. By taking into account both transmission losses and integration density, we choose parameters of $p = 0.5$, $R_{min} = 150\,\mu m$, and $R_{eff} = 220\,\mu m$. The transmission loss of fundamental TE mode for a single partial Euler bend is calculated as 0.0012 dB.

The 8-meter-long OTDL was fabricated on a commercially available 4-inch, 500 nm-thick X-cut TFLN wafer (NANOLN) using the standard PLACE technique. The fabrication process is summarized as follows: First, a 200 nm thick Cr film is deposited onto the surface of the TFLN wafer using magnetron sputtering. Then, the Cr film is patterned using femtosecond laser ablation. Next, the entire wafer undergoes a CMP process, during which the LN in areas not protected by the chromium mask is selectively etched, resulting in exceptionally smooth sidewalls for the photonic structure. The CMP etching depth is set as 210 nm, and the etching depth variation across the entire 4-inch wafer is less than 5 nm thanks for the mature CMP technique. After this, the Cr mask is removed by wet etching, and the wafer undergoes a secondary CMP process to smooth the upper surface for the LN waveguides. Then, the entire wafer undergoes an RCA clean, which is crucial because even the slightest contamination can lead to significant losses in the device. After the cleaning process, a 1.5 μm thick SiO2 cladding was deposited on the surface of the wafer using plasma-enhanced chemical vapor deposition (PECVD). This layer protects the LN waveguides from contamination during the subsequent fabrication and testing. Finally, the sample was sliced from the wafer, and both end facets were polished to minimize fiber coupling loss. Figure 2 (a) displays a photograph of the OTDL, taken with a digital camera. Figure 2 (b) presents a close-up view of the OTDL captured using an optical microscope. Figure 2 (c) displays a cross-sectional image of the waveguide with a scanning electron microscope (SEM). Figure 2 (d) shows the numerically simulated optical field profile of the fundamental TE mode at a wavelength of 1550 nm in the TFLN waveguide.

To evaluate the optical transmission loss of the fabricated OTDL, we measured its transmission spectrum over a wide spectral range using the setup depicted in Figure 3 (a). The light was generated from a tunable laser with a tunable wavelength ranging from 1510 nm to 1630 nm. An in-line polarization controller was used to adjust the light to TE polarization. The light was coupled into and out of the sample using a pair of ultra-high numerical aperture (UHNA) fiber arrays and fed into a broadband photodetector. We measured the loss of seven straight waveguides fabricated on the same chip to assess the coupling loss between the fiber and the waveguides. The results are shown in Figure 3 (b), indicating an average total coupling loss of 15.9 dB with a standard deviation of 1.7 dB. Given the short length (3 cm) of the straight waveguides, their intrinsic transmission loss was negligible. Figure 3 (c) shows the optical transmission loss derived from the transmission spectrum of the 8-meter-long OTDL after removing the coupling loss. The error from the coupling efficiency fluctuation is calculated as 0.002 dB/cm and is represented by the blue regions in the figure. As shown in Figure 3 (c), the average optical transmission loss of the fabricated OTDL is 0.036 dB/cm in the conventional band (1530 nm to 1565 nm) and 0.040 dB/cm over the entire range of 1510 nm to 1630 nm. The transmission loss is as low as 0.025 dB/cm near 1550 nm and 1580 nm. The relatively high losses observed near 1530 nm, 1560 nm and 1590 nm are attributed to absorption in the $SiO_2$ cladding, which could be reduced by optimizing the PECVD process parameters in future work.

Data buffering is a typical application of OTDLs in microwave photonics. To evaluate the performance of the fabricated OTDL as a data buffer, we transmitted a 0.5 GSPS non-return-to-zero (NRZ) data stream through the TFLN OTDL and measured the eye diagram of the output signal. The experimental setup is depicted in Figure 4 (a). A 1550 nm laser was modulated using an external electro-optic modulator and then coupled into and out of the fabricated OTDL. After amplification, the optical signal was injected into a high-speed photodetector, where the optical signal was converted into an electrical signal and subsequently captured by an oscilloscope to record the eye diagram. Figure 4 (b) displays the eye diagram of the signal taken directly from the modulator output, while Figure 4 (c) shows the eye diagram measured after transmission through the OTDL. As indicated, although there is an increase in noise due to reduced optical power, the OTDL did not introduce any additional distortion to the signal.

## 5. Conclusion and Outlook

In summary, we successfully demonstrate the fabrication and characterization of an 8-meter-long OTDL based on the TFLN platform. Thanks to the ability to produce highly uniform and low-loss photonic devices with PLACE technique, as well as the optimized low-loss, low-crosstalk, and compact partial Euler waveguide bends, our device achieves an average transmission loss as low as 0.036 dB/cm within the telecom C band. Given its length, which significantly exceeds that of previously reported TFLN photonic devices, this work highlights the impressive potential of the PLACE process for fabricating future ultra-large-scale TFLN devices.

## Acknowledgements

Y.R. and Y.N. contributed equally to this work. The work is supported by the National Key R&D Program of China (2019YFA0705000, 2022YFA1404600, 2022YFA1205100), National Natural Science Foundation of China (Grant Nos. 12334014, 12192251, 12134001, 12304418, 12274130, 12004116, 12274133), Science and Technology Commission of Shanghai Municipality (NO.21DZ1101500), Shanghai Municipal Science and Technology Major Project (Grant No.2019SHZDZX01), Innovation Program for Quantum Science and Technology (2021ZD0301403).

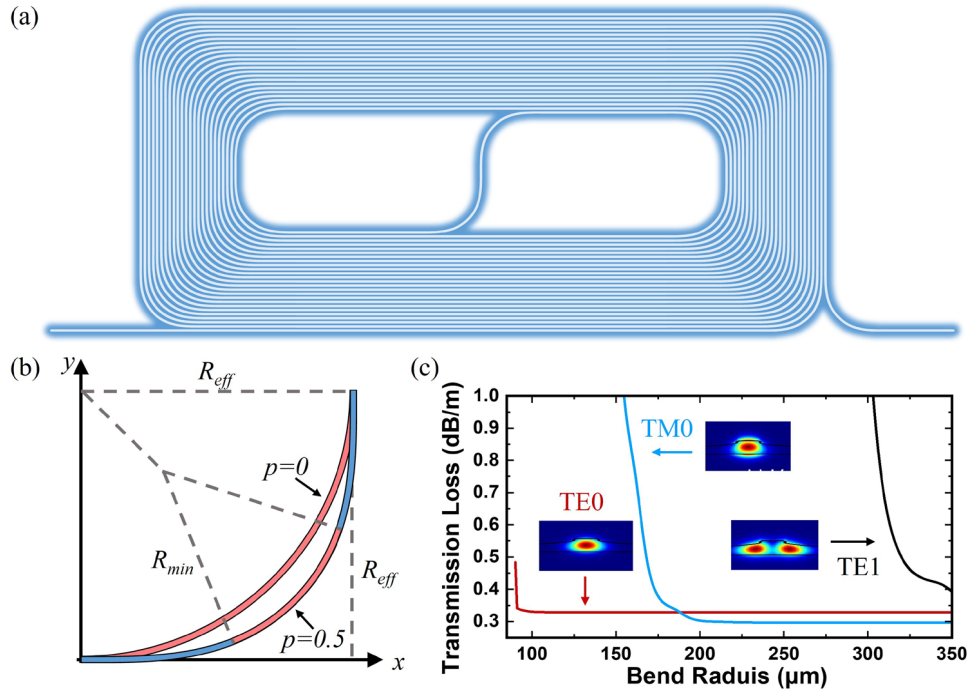

**Figure 1.** (a) Schematic of the OTDL demonstrated in this study. (b) Geometry of a 90° partial Euler waveguide bend, showing the curvature-continuous Euler (blue) and circular arc (red) regions. (c) Numerically simulated transmission losses of TFLN waveguides with different curvatures.

(a)

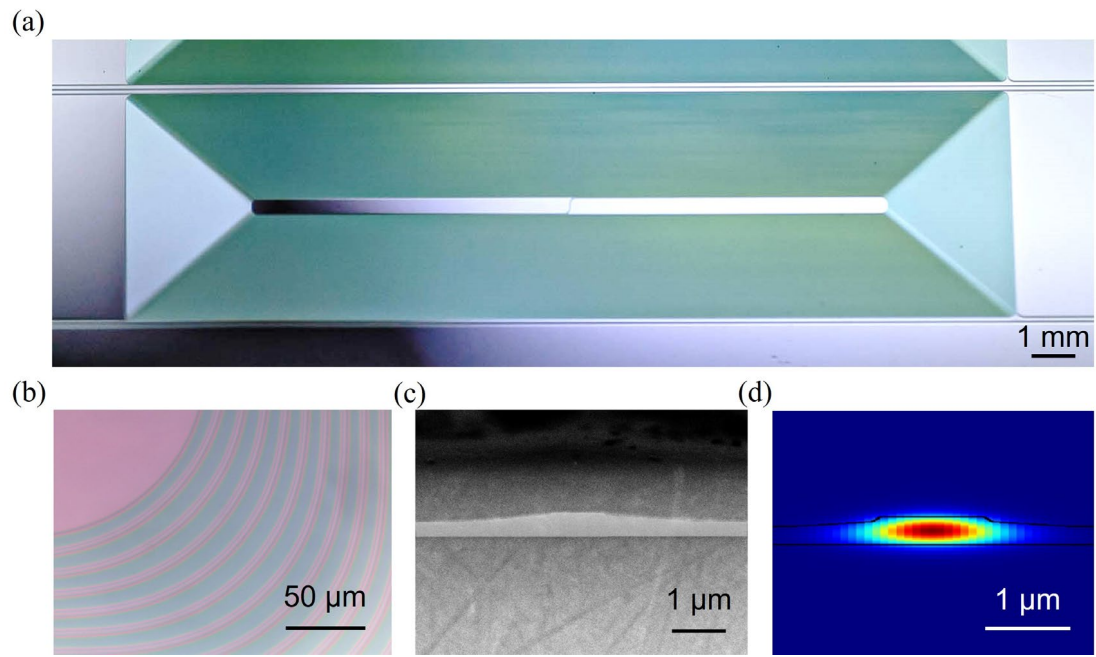

(b)

(c)

(d)

**Figure 2.** (a) Photograph of the OTDL captured with a digital camera. (b) Close-up view of the OTDL taken using an optical microscope. (c) Cross-sectional image of a waveguide captured with a scanning electron microscope (SEM). (d) Numerically simulated optical field profile of the fundamental TE mode at a wavelength of 1550 nm in the waveguide.

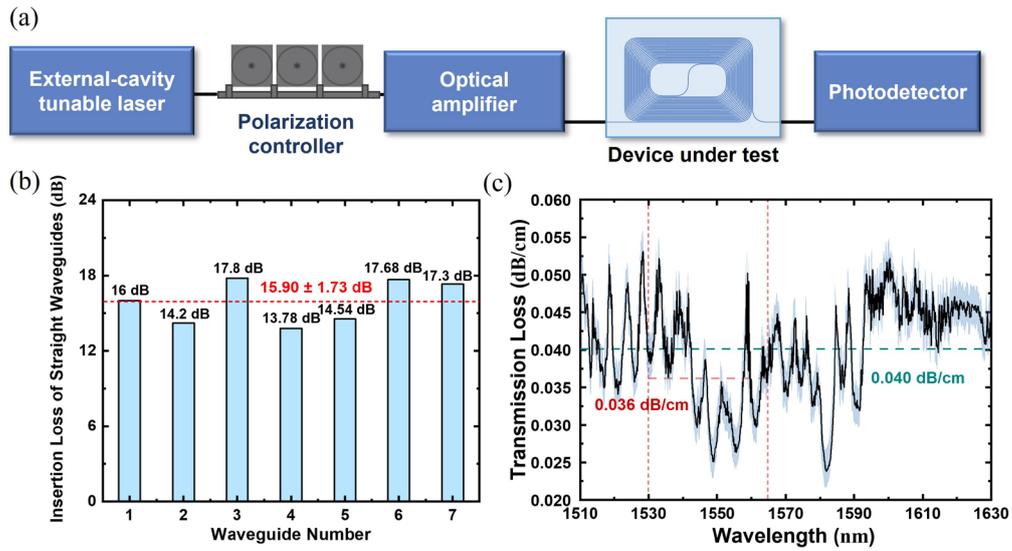

**Figure 3.** (a) Schematic of the experimental setup used to measure the optical transmission loss of the OTDL. (b) Measured coupling loss of seven straight waveguides on the same chip, showing an average total coupling loss of 15.9 dB with a standard deviation of 1.73 dB. (c) Optical transmission loss of the 8-meter-long OTDL derived from its transmission spectrum after removing coupling loss. The blue regions indicate coupling error.

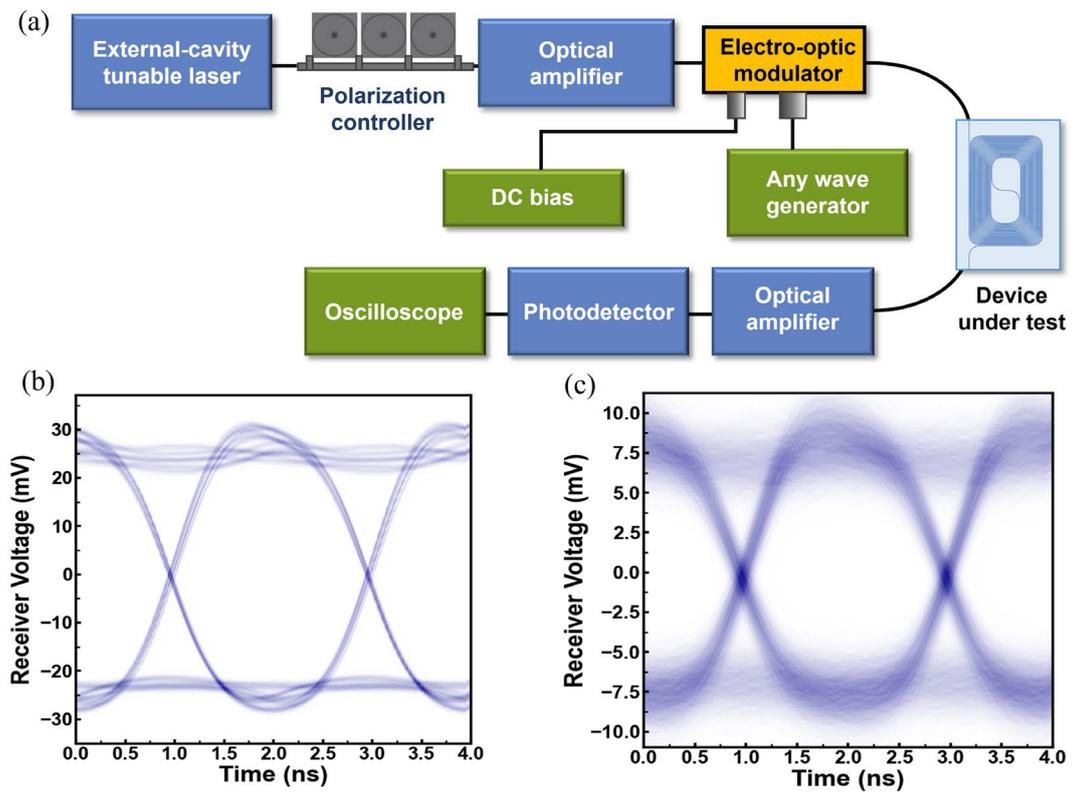

**Figure 4.** (a) Schematic of the experimental setup for testing the data buffering performance of the OTDL. (b) Eye diagram of the signal directly from the modulator output. (c) Eye diagram of the signal after transmission through the OTDL.